\documentclass[11pt]{article}
\usepackage{moriond,epsfig}

\def\beq{\begin{equation}}
\def\eeq{\end{equation}}
\def\bea{\begin{eqnarray}}
\def\eea{\end{eqnarray}}

\def \cn{Collaboration}
\def \ite{{\it et al.}}
\begin{document}
\rightline{EFI 03-17}
\rightline{hep-ph/0304240}
\bigskip
\centerline{\it Presented at the XXXVIII Rencontre de Moriond:
Electroweak Interactions}
\centerline{\it and Unified Theories, Les Arcs, France, March 15--22, 2003}

%\def\CPbar{\hbox{{\rm CP}\hskip-1.80em{/}}}
%temp replacement due to no font
%%%%%%%%%%%%%%%%%%%%%%%%%%%%%%%%%%%%%%%%%%%%%%%%%%
%                                                %
%    BEGINNING OF TEXT                           %
%                                                %
%%%%%%%%%%%%%%%%%%%%%%%%%%%%%%%%%%%%%%%%%%%%%%%%%%
%\vspace*{4cm}
\vspace*{2cm}
\title{P AND CP VIOLATION IN $B$ DECAYS}

\author{Michael Gronau\footnote{Permanent Address: Physics
Department, Technion -- Israel Institute of Technology, 32000 Haifa,
Israel.}}

\address{Enrico Fermi Institute and Department of Physics,\\
University of Chicago, Chicago, Illinois 60637}

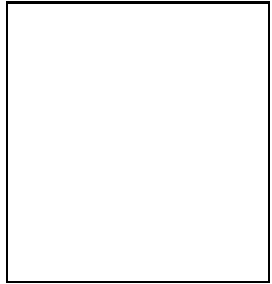
\maketitle\abstracts{
While the Kobayashi--Maskawa model of CP violation
passed its first crucial precision test in $B\to J/\psi K_S$,
other CP asymmetries in $B$ and $B_s$ decays will have to be measured
in order to critically test and overconstrain the model in
an unambiguous way. On another front, the chirality of weak $b$-quark
couplings has not yet been carefully tested. We discuss recent proposals
for studying both the chiral and CP-violating phase structures of these
couplings in $B\to D^*\rho$ and $B\to D^*a_1$.
}

\section{Introduction}
The Kobayashi--Maskawa model for CP violation was suggested
thirty years ago~\cite{KM} to explain the tiny CP non-conservation
observed in $K$ decays. In the past two years the model passed in a
remarkable way its first crucial test in $B$ decays~\cite{sanda} when
a large CP asymmetry was measured~\cite{psiKS} in $B^0 \to J/\psi K_S$,
in excellent agreement with expectations. The great virtue of this
decay mode is the absence of hadronic uncertainties~\cite{gold} in
predicting the {\em mixing induced asymmetry} in terms
of a fundamental phase parameter $\beta\equiv \phi_1$ of the Standard
Model. This opens a new era, in which other CP asymmetries in $B$ and
$B_s$ decays will have to be measured in order to test the weak phase
structure of $b$ quark couplings in an unambiguous way. Hopefully, this
will lead to a point where deviations from the simple CKM framework will
be observed. An important step in this direction would be a measurement
of the phase $\gamma\equiv \phi_3$, usually associated with {\em CP
violation in direct decays}. Several methods along this line, in which
experimental progress has been made recently, were discussed at this
meeting.\,\cite{Rosner,Nakadaira,Hamel,Tomura,Swain}

The most accessible experimental tests for $\gamma$, in processes such as
$B~(B_s)\to~\pi\pi$~$(KK)$ and $B~(B_s) \to K\pi$, involve theoretical
hadronic uncertainties due to penguin amplitudes, which may be
resolved by applying approximate symmetries such as isospin or flavor
SU(3) including SU(3) breaking effects.\,\cite{Rosner} Other tests,
including $B^\pm\to DK^\pm$,\,\cite{Swain,GW} which
are free of such uncertainties, usually require a larger number of $B$
mesons than produced so far. Our present discussion will focus on two
theoretically clean studies of $B^0 \to D^-\rho^+$ and $B^0\to D^-a^+_1$,
where time-dependent CP asymmetries are sensitive to the weak phase
$2\beta + \gamma$, combining the mixing phase $2\beta$ and the decay phase
$\gamma$. While these studies are very challenging from an experimental
point of view, and require much more data than accumulated so far, we will
explain first what can be learned from existing data of these processes,
without performing time-dependent measurements. In particular, we will
show how to test the left-handed chirality of the weak $b$ quark coupling,
for which very little evidence exists.\,\cite{MG}

Our motivation for calling for chirality tests of $b$ quark coupling is
both phenomenological (1) as well as theoretical (2):
\begin{enumerate}
\item The very small charged current $b$ quark couplings,\,\cite{PDG}
$|V_{cb}| = 0.04,~|V_{ub}| = 0.003-0.004$, are more sensitive to new
types of interactions, such as right-handed currents, than the lighter
quarks' couplings.
\item An artificial left-right asymmetry is introduced by hand in the
Standard Model in order to account for the low energy weak interaction
phenomenology. Ultimately, left-right symmetry may be restored at high
energies.\,\cite{LR} If parity violation and the observed quark mass
hierarchy (also introduced by hand in the Standard Model in terms of
arbitrary Yukawa couplings) have a common origin, then it may be
expected that right-handed couplings grow with quark masses and are
larger for the $b$ quark than for $s$ and $d$ quarks.
\end{enumerate}

Our discussion will start with chirality tests for the $b$ coupling and will
end with studies of CP violation.
In Section 2 we study helicity amplitudes in $B\to D^*\rho$, pointing
out the success of predicting these amplitudes using
factorization and heavy quark symmetry. Application of the same assumptions
to $B \to D^*a_1$ is shown in Section 3 to permit a test of V--A for the
$b$ quark coupling. In Section 4 time-dependent CP asymmetries
in $B\to D^*\rho$ and $B \to D^*a_1$ are studied in order
to learn $2\beta +\gamma$, while Section 5 concludes.

\section{The decay $\bar B^0 \to D^{*+}\rho^-$}
\subsection{Helicity amplitudes}
The decays $\bar B^0 \to D^{*+}~(\to D^0\pi^+)~\rho^-~(\to \pi^-\pi^0)$,
in which each of the two vector mesons decays to two spinless particles
whose momenta are measured, can be used to study the vector meson
polarization.\,\cite{KG} Using an angular momentum decomposition, the decay
amplitude can be written in terms of
three helicity amplitudes, $H_0,~H_+,~H_-$, corresponding to the three
polarization states of the vector mesons,
\beq\label{A}
A = \frac{3}{2\sqrt {2\pi}}\left [H_0\cos\theta_1\cos\theta_2 +
\frac{1}{2}(H_+ e^{i\phi} + H_- e^{-i\phi})\sin\theta_1\sin\theta_2\right ]~~.
\eeq
Here $\theta_1$ and $\theta_2$ are the angles between each of the two vector
mesons' momenta in the $B$ rest frame and the momenta of the corresponding
daughter particles in the decaying vector mesons' rest frame; $\phi$ is
the angle between the $D^*$ and $\rho$ decay planes.
We use a convention in which the normalized decay angular
distribution is given by $|A|^2$,
\beq
\frac{1}{\Gamma}\frac{d^3\Gamma}{d\cos\theta_1\cos\theta_2 d\phi} =
|A|^2 ~~\Rightarrow~~
|H_0|^2 + |H_+|^2 + |H_-|^2 = 1~~.
\eeq

The decay distribution is symmetric under $(H_0,~H_+,~H_-) \to
(H^*_0,~H^*_-,~H^*_+)$, implying that rates into left and right
polarizations, $|H_-|^2$ and $|H_+|^2$ respectively,
are indistinguishable. Namely, one cannot distinguish in this process
between left- and right-polarized vector mesons. As will be explained in
the next section, this follows from the lack of a parity-odd observable
when each of the vector mesons decays into two spinless particles.
Thus, while the rate into the longitudinally polarized state $|H_0|^2$
can be  measured, only the magnitude of $|H_+|^2 - |H_-|^2$ is measurable,
but not its sign.

\subsection{Factorization and heavy quark symmetry}
The three helicity amplitudes $H_{0,\pm}$ can be calculated using
factorization and heavy quark symmetry.\,\cite{BBNS2}
Factorization implies
\beq
\langle D^*\rho | H_{\rm eff}| \bar B\rangle \propto
\langle D^* |V_\mu - A_\mu | \bar B\rangle \langle \rho |V^\mu |0\rangle~~,
\eeq
where $\langle \rho(\epsilon) |V^\mu |0\rangle  \propto  \epsilon^\mu$.
In the heavy quark symmetry limit the $V-A$ current matrix element can
be written in terms of a single form factor multiplying a purely kinematic
factor,
\beq
\langle D^*(v', \epsilon') |V_\mu - A_\mu| \bar B(v)\rangle  \propto
i\epsilon_{\mu\nu\alpha\beta}\epsilon'^{\nu}v^\alpha v'^\beta +
\epsilon'_\mu(1 + v\cdot v') - v'_\mu\epsilon'\cdot v~~,
\eeq
where $v\equiv p/m$. In this approximation, the three normalized helicity
amplitudes can be written in terms of meson
masses. For a $\bar c \gamma_\mu (1 - \gamma_5) b$ current one
finds~\cite{factor}
\beq\label{H}
H_0 = \left (1 + \frac{4y}{y + 1}\epsilon^2\right )^{-\frac{1}{2}}~~,~~~~~
H_\pm = \left ( 1 \mp \sqrt{\frac{y-1}{y+1}}\right )\epsilon
\left (1 + \frac{4y}{y + 1}\epsilon^2\right )^{-\frac{1}{2}}~~,
\eeq
where $y \equiv (m_B^2+m_{D^*}^2-m_{\rho}^2)/2m_B m_{D^*} = 1.476,~\epsilon
\equiv
m_\rho/(m_B - m_{D^*}) = 0.236$. Thus, one obtains the values
\beq\label{Hi}
H_0 = 0.940~,~~~~ H_+ = 0.125~,~~~~H_- = 0.318~~.
\eeq
These predictions of the Standard Model apply to $\bar B^0$ decays,
while in $B^0$ decays the values of $H_+$ and $H_-$ are
interchanged. In the case of a $\bar c \gamma_\mu (1 + \gamma_5) b$
current, the roles of $H_+$ and $H_-$ are interchanged.

The following values were reported very recently by the CLEO collaboration
for $\bar B^0 \to D^{*+}\rho^-$:\,\cite{CLEO}
\bea\label{CLEO}
|H_0| & = & 0.941 \pm 0.009 \pm 0.006~~,\nonumber \\
|H_+|~{\rm or}~|H_-| & = & 0.107 \pm 0.031 \pm 0.011~~,\nonumber \\
|H_-|~{\rm or}~|H_+| & = & 0.322 \pm 0.025 \pm 0.016~~.
\eea
The collaboration quotes a value for $|H_+|$ which is
smaller than $|H_-|$, assuming that the $D^*$ predominantly carries
the chirality of the $c$ quark, as would follow from a
$\bar c\gamma_\mu(1 - \gamma_5)b$ coupling. While it is impossible to
check this assumption in this experiment, it is important to verify it
elsewhere.

The experimental results (\ref{CLEO}) are in very good agreement with
the amplitudes calculated in Eq.~(\ref{Hi}). In particular, the measured
value of $|H_0|$ agrees with the Standard Model prediction at
a very high precision.
The agreement between theory and experiment for $|H_{\pm}|$ is
somewhat surprising, since one expects sizable deviations from
factorization in these amplitudes which are by themselves
subleading in $1/m_b$. In principle, order $1/m_b$ corrections to
factorization in these amplitudes could be as large as the amplitudes
themselves.
There are some hints in the data~\cite{CLEO} for possibly nonzero relative
final state interaction phases between $H_{\pm}$ and $H_0$, which would
indicate deviations from factorization.
We note that, in spite of this outstanding agreement
between factorization predictions and
experiment, measurements of $|H_{\pm}|$ cannot distinguish
between $V-A$ and $V+A$ currents. The present experimental precision in
the measured values of $|H_{\pm}|$ allows also for an admixture of the two
opposite chiralities in the $b \to c$ coupling.

\section{The decay $\bar B^0 \to D^{*+}a^-_1$}
\subsection{What's unique about $B\to D^*a_1$?}
In the previous Section we noted that in $B$ meson decays to two vector
mesons, each of which decays to two spinless particles whose momenta are
measured, one cannot distinguish between left- and right-polarized vector
mesons. Therefore, these processes are unsuitable for chirality tests for
the $b$ quark coupling. A chirality measurement for one of the two decay
particles requires that the particle decays subsequently to a three body
final state.
The sequence of arguments proving this statement is straightforward:
(a) Chirality is a parity-odd quantity, (b) Hadronic quantities
multiplying the chirality in the decay distribution must be parity-odd,
(c) A pseudoscalar quantity containing the smallest number of hadron
momenta is a triple product, $\vec p_1\cdot (\vec p_2\times\vec p_3)$.
Thus, chirality measurements can be performed in $B$ meson decays into a
vector meson and an axial-vector meson, which decays subsequently to
three pseudoscalars. We note in passing that decays into two vector mesons,
one of which decays to three pseudoscalars, do not contain sufficient
invariants to permit such a measurement.\,\cite{GGPR}

This leads us to the decay $\bar B^0\to D^{*+}a^-_1$,
in which the $a_1$ is observed through $a_1^- \to \pi^-\pi^-\pi^+$.
This decay mode is unique in the following sense.\,\cite{GPW}
A triple product $\vec p_1\cdot (\vec p_2\times\vec p_3)$ is not only
parity-odd but also time-reversal-odd, which requires a nonzero phase due
to final state interactions. Usually, such a phase is incalculable and would
render measurements which cannot be interpreted theoretically in a simple
manner. In the case of $a_1  \to \pi^-\pi^-\pi^+$, the decay occurs
through an interference of two intermediate $\rho^0$ states, the
amplitudes of which are equal by isospin. Thus, the final state phase
is calculable in terms of the $\rho$ width.

\subsection{Distinguishing between left-handed and right-handed helicities}
The decay amplitude for $\bar B^0\to D^{*+}a_1^-,~
a_1^-\to \pi^-\pi^-\pi^+$ is written in terms of weak helicity amplitudes
$H'_i$, in analogy with (\ref{A}),
\beq
A(\bar B^0\to D^{*+} \pi^-(p_1)\pi^-(p_2)\pi^+(p_3)) = \sum_{i=0,+,-}
H'_i A_i~~.
\eeq
The strong amplitude $A_i$ involves two terms, corresponding to two
possible ways
of forming a $\rho$ meson from $\pi^+\pi^-$ pairs, each of which can be
written
in terms of two invariant amplitudes:
\beq\label{a1rhopi}
A(a_1(p,\varepsilon)\to \rho(p',\varepsilon') \pi) =
A(\varepsilon\cdot \varepsilon'^*) + B(\varepsilon\cdot p')
(\varepsilon'^*\cdot p)~,
\eeq
convoluted with the amplitude for $\rho^0(\varepsilon') \to
\pi^+(p_i)\pi^-(p_j)$, which
is proportional to $\varepsilon'\cdot(p_i - p_j)$.
One finds \cite{GPW}
\bea\label{a1amp}
& & A(a_1^-(p,\varepsilon) \to \pi^-(p_1) \pi^-(p_2) \pi^+(p_3)) \propto
C(s_{13}, s_{23}) (\varepsilon\cdot p_1) +(p_1 \leftrightarrow p_2)~,\\
\label{C}
& & C(s_{13}, s_{23}) = [A + B m_{a_1} (E_3-E_2)] B_\rho (s_{23}) +
2A B_\rho (s_{13})~~,
\eea
where $s_{ij} = (p_i+p_j)^2,~B_{\rho}(s_{ij}) = (s_{ij} - m^2_{\rho} -
im_{\rho}
\Gamma_{\rho})^{-1}$, and pion energies are given in the $a_1$ rest frame.
The amplitudes
$A$ and $B$ are related to $S$- and $D$-wave $\rho\pi$ amplitudes.
When neglecting the small $D$-wave amplitude,\,\cite{PDG} they
obey~\cite{SD}
\beq\label{AB}
B = -A \left(1-\frac{m_\rho}{E_\rho}\right)\frac{E_\rho}{m_\rho \vec
p_\rho\,^2}~~.
\eeq

Defining an angle $\theta$ between the normal to the $a_1$ decay plane,
$\hat n$, and the direction opposite to the $D^*$ in the $a_1$ rest frame,
one calculates the $B\to D^* 3\pi$ decay distribution,
\bea\label{dist}
\frac{d\Gamma}{ds_{13}ds_{23}d\cos\theta} & \propto &
|H'_0|^2 \sin^2\theta |\vec J|^2 +
(|H'_+|^2 + |H'_-|^2)\frac12 (1 + \cos^2 \theta) |\vec J|^2\nonumber \\
& + & (|H'_+|^2 - |H'_-|^2) \cos\theta\,\mbox{Im}[(\vec J\times
\vec J^*)\cdot \hat n]~~,
\eea
where
\beq\label{J}
\vec J = C(s_{13}, s_{23}) \vec p_1 + C(s_{23}, s_{13}) \vec p_2~~.
\eeq
A fit to the angular decay distribution enables separate measurements
of the three terms $|H'_0|^2,~|H'_+|^2 + |H'_-|^2$ and $|H'_+|^2 -
|H'_-|^2$.
We note that when calculating the quantity $\vec J$ {\em free of any
parameter}, using Eq.~(\ref{AB}), we have only assumed for the $a_1$
an $S$-wave $\rho^0\pi^-$ structure, without using the $a_1$ resonance
shape and width, which would have involved a large uncertainty.\,\cite{PDG}
A small $D$-wave correction can also be incorporated in the calculation.

\subsection{Factorization and a chirality test}
In the heavy quark symmetry and factorization
approximation,\,\cite{BBNS2,factor} using
(\ref{H}) where $y \equiv (m_B^2+m_{D^*}^2-m_{a_1}^2)/2m_B m_{D^*} = 1.432,~
\epsilon \equiv m_{a_1}/(m_B - m_{D^*}) = 0.376$ , the results for a
$\bar c\gamma_\mu(1-\gamma_5)b$ current are
\beq\label{H'i}
H'_0 = 0.866~,~~~~ H'_+ = 0.188~,~~~~H'_- = 0.463~.
\eeq
These values, which depend somewhat on $m_{a_1}$, can be verified by
measuring the decay distribution (\ref{dist}).

In order to define a measure for the sensitivity of determining the
chirality of the $b$ quark coupling,\,\cite{better} let us consider the
following P-odd up--down asymmetry of the $D^*$ momentum direction with
respect to the
$a_1$ decay plane:
\beq
{\cal A} = \frac{\int_0^{\pi/2}\frac{d\Gamma}{d\theta}d\theta -
\int_{\pi/2}^\pi \frac{d\Gamma}{d\theta}d\theta}{\Gamma}~~.
\eeq
One has
\beq\label{As}
{\cal A} = \frac{3}{4} \frac{\langle \mbox{Im }[\hat n\cdot (\vec J\times
\vec J^*)]
\mbox{sgn }(s_{13}-s_{23})\rangle}{\langle |\vec J|^2\rangle}
\frac{|H'_+|^2 - |H'_-|^2}{|H'_0|^2+|H'_+|^2 + |H'_-|^2}~~,
\eeq
and  integration over the entire Dalitz plot gives
\beq
{\cal A} = -0.237 \frac{|H'_+|^2 - |H'_-|^2}{|H'_0|^2+|H'_+|^2 + |H'_-|^2}~~.
\eeq
Measuring this asymmetry determines $|H'_+|^2 - |H'_-|^2$.
Using Eq.~(\ref{H'i}), one obtains ${\cal A} = 0.042$.

The sign of the asymmetry provides an unambiguous signature for
a $V-A$ coupling, in contrast to $V+A$ which would yield an opposite sign.
In the center-of-mass frame of $\pi^-\pi^-\pi^+$ {\em the $\bar B^0$ and
$D^{*+}$ prefer to move in the hemisphere defined by the direction
$\vec{p}(\pi^-)_{\rm fast}\times \vec{p}(\pi^-)_{\rm slow}$}. In order
to measure an asymmetry at this level one needs about 5000 identified
$B \to D^*a_1$ events. A very large sample of $18400 \pm 1200$ partially
reconstructed events was reported recently
by the BaBar collaboration,\,\cite{BABARa1}
in which the $a_1$ was reconstructed via the decay
chain $a^-_1 \to \rho^0\pi^-,~\rho^0\to \pi^+\pi^-$ while the $D^*$ was
identified by a slow pion. A correspondingly smaller
sample of fully reconstructed events seems sufficient for an up-down
asymmetry measurement.
A more precise measurement of $|H'_+|^2 - |H'_-|^2$ than from the
asymmetry alone may be obtained by fitting data to the energy- and
angle-dependent decay distribution given in Eq.~(\ref{dist}).

\section{Determining $2\beta + \gamma$ in time-dependent decays}
Both $\bar B^0 \to D^{*+}\rho^-$ and $\bar B^0 \to D^{*+}a^-_1$ belong to
a class of processes, which also contains $\bar B^0 \to D^+\pi^-,
D^{*+}\pi^-,~D^+\rho^-$,\,\cite{ID} from which the weak phase
$2\beta + \gamma$ can be determined with no hadronic
uncertainty. Using the well-measured value \cite{psiKS} of $\beta$
this would fix $\gamma$. The difficulty in these methods
lies in having to measure a very small time-dependent
interference between $b\to c \bar u d$ and doubly-CKM-suppressed $\bar b
\to \bar u c \bar d$ transitions, where $|V^*_{ub}V_{cd}/V_{cb}V^*_{ud}|
\simeq 0.02$.

In decays to $D^+\pi^-,~D^{*+}\pi^-,~D^+\rho^-$ the resulting
analyses are sensitive to the square of a doubly-CKM-suppressed
amplitude, a precise knowledge of which is very challenging.
In decays to two vector mesons, $\bar B^0 \to D^{*+}\rho^-$,
one avoids the need to determine this small quantity by using an
interference between helicity amplitudes of CKM-allowed
and doubly-CKM-suppressed decays. This was claimed~\cite{LSS} to
improve the sensitivity, but requires a detailed angular analysis in
addition to time-dependent measurements. The feasibility of using an
angular analysis for measuring the helicity amplitudes in the dominant
CKM-allowed channel was demonstrated by the CLEO collaboration~\cite{CLEO}
as discussed above.
It will require considerably more statistics to measure the time and
angular dependent interference of helicity amplitudes with such disparate
magnitudes. Here we will assume that sufficient statistics
is gained,\,\cite{Wilson}
and will describe this method for determining $2\beta + \gamma$, first
in $\bar B^0 \to D^{*+}\rho^-$,\,\cite{LSS} and then in $\bar B^0 \to
D^{*+}a^-_1$,\,\cite{GPW} where a discrete ambiguity in the weak phase
will be shown to be resolved.

\subsection{$\bar B^0(t) \to D^{*+}\rho^-$}
It is convenient to write the amplitude $A \equiv A(\bar B^0 \to
D^{*+}(\to D^0\pi^+)
\rho^-(\to\pi^-\pi^0))$ in a linear polarization basis (a so-called
transversity basis~\cite{factor,DQSTL}), in which the
$D^*$ and $\rho$ transverse polarizations are either parallel or
perpendicular to one another,
$H_{\parallel,\perp} = (H_+ \pm H_-)/\sqrt{2}$, and to similarly expand
$a \equiv A(B^0 \to D^{*+}\rho^-)$ in terms of $h_{0,\parallel,\perp}$:
\bea\label{Aa}
A & = & \frac{3}{2\sqrt {2\pi}}(H_0g_0 + H_\parallel g_\parallel +
iH_\perp g_\perp)~,~~
a = \frac{3}{2\sqrt {2\pi}}(h_0g_0 + h_\parallel g_\parallel +
ih_\perp g_\perp)~,\\
\label{gt}
g_0 & = & \cos\theta_1\cos\theta_2~,~g_\parallel =
\frac{1}{\sqrt 2}\sin\theta_1\sin\theta_2\cos\phi~,~g_\perp =
\frac{1}{\sqrt 2}\sin\theta_1\sin\theta_2\sin\phi~~.
\eea
The transversity amplitudes can be written as
\beq
H_t = |H_t|\exp(i\Delta_t)~~,~~~~~~h_t = |h_t|\exp(i\delta_t)\exp(i\gamma)~~.
\eeq

The time-dependent rate for $\bar B^0(t) \to D^{*+}\rho^-$ has the
general form
\bea\label{Bt}
\Gamma(t) & \propto & e^{-\Gamma t}\left [(|A|^2 + |a|^2) + (|A|^2 -
|a|^2)\cos\Delta mt\right .
+  \left. 2 {\rm Im}\left( e^{-2i\beta} A a^*\right)\sin \Delta
mt\right ]\nonumber\\
& = & e^{-\Gamma t}\sum_{t \le t'}\left (\Lambda_{tt'} +
\Sigma_{tt'}\cos\Delta mt
+ \rho_{tt'}\sin\Delta mt \right )g_tg_{t'}~.
\eea
Each of the coefficients in the sum can be measured by performing a
time-dependent
angular analysis. Denoting $\Phi \equiv 2\beta + \gamma$, this
determines the following quantities:
\bea\label{terms}
& & |H_t|^2~,~~~|H_0||H_\perp|\sin(\Delta_0 - \Delta_\perp)~,~~~
|H_\parallel||H_\perp|\sin(\Delta_\parallel -
\Delta_\perp)~~,\nonumber \\
& & |H_t||h_t|\sin(\Phi +\Delta_t -\delta_t )~,~~~~~~~~~~~~~~~~~~~~~~~~~~~~~
(t = 0,\parallel,\perp )~~,
\nonumber \\
& & |H_\perp|h_0|\cos(\Phi  + \Delta_\perp - \delta_0) -
|H_0|h_\perp|\cos(\Phi  + \Delta_0 - \delta_\perp)~~,\nonumber \\
& & |H_\perp|h_\parallel|\cos(\Phi  + \Delta_\perp - \delta_\parallel) -
|H_\parallel|h_\perp|\cos(\Phi  + \Delta_\parallel - \delta_\perp)~~.
\eea
One does not rely on knowledge of the small $|h_t|^2$ terms,\,\cite{LSS}
in which uncertainties would
be large. Decays into the charge-conjugate state $D^{*-}\rho^+$
determine similar quantities, where $\Phi$ is replaced by $-\Phi$. It is
then straightforward to show that
this overall information is sufficient for determining $|\sin\Phi|$.
However, the sign of $\sin(2\beta +\gamma)$ remains ambiguous.

\subsection{What is new in the time-dependence of $\bar B^0(t)
\to D^{*+}a^-_1$ ?}
The amplitudes $A'\equiv A(\bar B^0 \to D^{*+}(3\pi)^-_{a_1})$ and $a'
\equiv A(B^0 \to
D^{*+}(3\pi)^-_{a_1})$ are written in analogy with (\ref{Aa}):
\bea
A' = \sum_{t=0,\parallel,\perp} H'_t A_t~~,~~~~
a' = \sum_{t=0,\parallel,\perp} h'_t A_t~~.
\eea
Instead of the {\em real} functions $g_t$ in Eq.~(\ref{gt}) of the angular
variables $\theta_1,~\theta_2$ and $\phi$, one has calculable {\em complex}
amplitudes $A_t$ defined in Eq.~(\ref{a1amp}). These are functions of
$\theta$ defined above, an angle $\chi$ describing a common angle of
rotation for the three pions in the $a_1$ decay plane, and an angle $\psi$
determined by the $D^*$ decay plane.\,\cite{GPW} The latter
defines the angle between the two intersection lines of the $D^*$ decay
plane and of the $a_1$ decay plane with a plane perpendicular to the $D^*$
direction.

One measures $\Gamma(\bar B^0(t) \to D^{*+}(3\pi)^-_{a_1})$
and $\Gamma(\bar B^0(t) \to D^{*-}(3\pi)^+_{a_1})$, with time-dependence
as in Eq.~(\ref{Bt}), as a function of $\theta$ and $\psi$ while
integrating over $\chi$.
Instead of the product of geometrical functions $g_tg_{t'}$ in $B\to D^*\rho$,
the sum in Eq.~(\ref{Bt}) now involves calculable functions of the angles
$\theta$ and $\psi$, defined as $R_{ij}\equiv (1/2\pi)\int\mbox{d} \chi
{\rm Re}(A_iA^*_j)$ and $I_{ij}\equiv (1/2\pi)\int\mbox{d} \chi
{\rm Im}(A_iA^*_j)$, ($i,j=0,\parallel,\perp$). The nine independent
functions are given by
\bea
R_{00} & = & \frac12\sin^2\theta |\vec J|^2~~,~~~~~~~~~~~~~~~~~~~~
R_{\parallel\,\parallel} = \frac12(1 - \cos^2\psi\sin^2\theta)
|\vec J|^2~,\nonumber \\
R_{\perp\perp} & = & \frac12(1 - \sin^2\psi\sin^2\theta)|\vec J|^2~~,~~~~~
R_{0\parallel} = \sin\psi\sin\theta J^2_n~, \nonumber \\
R_{0\perp} & = & \frac{1}{4}\sin\psi\sin 2\theta |\vec J|^2~~,~~~~~~~~~~~~~
R_{\parallel\perp} = \cos\theta J^2_n~, \nonumber \\
I_{0\parallel} & = & -\frac{1}{4}\cos\psi\sin 2\theta |\vec J|^2~~,~~~~~~~~~~~
I_{0\perp} = -\cos\psi\sin\theta J^2_n~, \nonumber \\
I_{\parallel\perp} & = & -\frac{1}{4}\sin 2\psi\sin^2\theta |\vec J|^2~~,
\eea
where $J^2_n \equiv (1/2)\mbox{Im}[(\vec J\times \vec J^*)\cdot \hat n]$.

The complex amplitudes $A_t$, in
contrast to the real functions $g_t$, imply
that one can measure both real and imaginary interference terms between
transversity amplitudes $H'_t$ and $h'_{t'}$. This includes terms similar to
those in Eq.~(\ref{terms}) in which the cosines and sines are interchanged.
These additional terms provide information which enables resolving the
ambiguity in the sign of $\sin(2\beta + \gamma)$.\,\cite{GPW}

The advantage of $B\to D^*a_1$ in determining {\em unambiguously} the
CP-violating phase
$2\beta + \gamma$ can be traced back to the parity-odd measurables
that occur
in this process but not in $B \to D^*\rho$. As noted, $|H'_+|^2 -
H'_-|^2 = 2{\rm Re}
(H'_\parallel H'^*_\perp)$ is P-odd, and so is ${\rm Im}[e^{2i\beta}
(H'_\parallel h'^*_\perp +
H_\perp h^*_\parallel )]$. These terms, which do not occur in the
time-dependent rate of
$\bar B^0 \to D^{*+}\rho^-$, do occur in $\bar B^0(t) \to D^{*+}a_1^-$
multiplying a P-odd function
of $\theta$, $\cos\theta\,\mbox{Im}[(\vec J\times \vec J^*)\cdot
\hat n]$. A practical advantage of
$\bar B^-\to D^{*+}a^-_1$ over $\bar B^0 \to D^{*+}\rho^-$ is the
occurrence of only charged pions
in the first process. A slight disadvantage of the first process may
be an intrinsic uncertainty in the amplitudes $A_t$ calculated in
Eq.~(\ref{a1amp}), due to a possible small D-wave $\rho\pi$ amplitude.

\section{Conclusion}
Predictions of factorization and heavy quark symmetry for helicity
amplitudes in $\bar B^0 \to D^{*-}\rho^+$ agree very well with
experiment, but  do not distinguish between positive and negative
helicities. Parity-odd measurables in hadronic $B$ decays are quite
rare. We identify such a measurable in $\bar B^0\to D^{*-}a_1^+$
in terms of the up-down asymmetry of the $D^*$ momentum direction with
respect to the $a_1$ decay plane. Measurement of this asymmetry using
current data can test the chirality of the weak $b$ quark coupling.
Time-dependent CP asymmetry measurements in $B \to D^*\rho$ and
$B \to D^* a_1$, which entail the potential for a clean determination
of $2\beta + \gamma$, require considerably more data than acquired
so far. Study of $\bar B^0(t) \to D^{*-} a_1^+$ complements that of
$\bar B^0(t) \to D^{*-}\rho^+$, and resolves a discrete ambiguity in
the CP-violating phase.

\section*{Acknowledgments}
I thank Dan Pirjol and Daniel Wyler for an enjoyable collaboration
on a study of $B\to D^*a_1$, and David London, Nita and Rahul Sinha
for pointing out the virtue of $B\to D^*\rho$.
I am grateful to the CERN Theory Division and the Enrico Fermi
Institute at the University of Chicago for their kind hospitality.
This work was supported in part by the United States Department of
Energy through Grant No.\ DE FG02 90ER40560.

% Journal definitions
\def \ajp#1#2#3{Am.\ J. Phys.\ {\bf#1}, #2 (#3)}
\def \apny#1#2#3{Ann.\ Phys.\ (N.Y.) {\bf#1}, #2 (#3)}
\def \app#1#2#3{Acta Phys.\ Polon. {\bf#1}, #2 (#3)}
\def \arnps#1#2#3{Ann.\ Rev.\ Nucl.\ Part.\ Sci.\ {\bf#1}, #2 (#3)}
\def \art{and references therein}
\def \cmts#1#2#3{Comments on Nucl.\ Part.\ Phys.\ {\bf#1}, #2 (#3)}
\def \cn{Collaboration}
\def \epjc#1#2#3{Eur.\ Phys.\ J.\ C. {\bf#1}, #2 (#3)}
\def \ib{{\it ibid.}~}
\def \ibj#1#2#3{~{\bf#1}, #2 (#3)}
\def \ijmpa#1#2#3{Int.\ J.\ Mod.\ Phys.\ A {\bf#1}, #2 (#3)}
\def \ite{{\it et al.}}
\def \jhep#1#2#3{JHEP {\bf#1}, #2 (#3)}
\def \jpb#1#2#3{J.\ Phys.\ B {\bf#1}, #2 (#3)}
\def \kdvs#1#2#3{{Kong.\ Danske Vid.\ Selsk., Matt-fys.\ Medd.} {\bf #1}, No.\
#2 (#3)}
\def \mpla#1#2#3{Mod.\ Phys.\ Lett.\ A {\bf#1}, #2 (#3)}
\def \nat#1#2#3{Nature {\bf#1}, #2 (#3)}
\def \nc#1#2#3{Nuovo Cim.\ {\bf#1}, #2 (#3)}
\def \nima#1#2#3{Nucl.\ Instr.\ Meth.\ A {\bf#1}, #2 (#3)}
\def \npb#1#2#3{Nucl.\ Phys.\ B~{\bf#1}, #2 (#3)}
\def \npps#1#2#3{Nucl.\ Phys.\ Proc.\ Suppl.\ {\bf#1}, #2 (#3)}
\def \os{XXX International Conference on High Energy Physics, Osaka, Japan,
July 27 -- August 2, 2000}
\def \PDG{Particle Data Group, D. E. Groom \ite, \epjc{15}{1}{2000}}
\def \pisma#1#2#3#4{Pis'ma Zh.\ Eksp.\ Teor.\ Fiz.\ {\bf#1}, #2 (#3) [JETP
Lett.\ {\bf#1}, #4 (#3)]}
\def \pl#1#2#3{Phys.\ Lett.\ {\bf#1}, #2 (#3)}
\def \pla#1#2#3{Phys.\ Lett.\ A {\bf#1}, #2 (#3)}
\def \plb#1#2#3{Phys.\ Lett.\ B {\bf#1}, #2 (#3)}
\def \prl#1#2#3{Phys.\ Rev.\ Lett.\ {\bf#1}, #2 (#3)}
\def \prd#1#2#3{Phys.\ Rev.\ D\ {\bf#1}, #2 (#3)}
\def \prp#1#2#3{Phys.\ Rep.\ {\bf#1}, #2 (#3)}
\def \ptp#1#2#3{Prog.\ Theor.\ Phys.\ {\bf#1}, #2 (#3)}
\def \rmp#1#2#3{Rev.\ Mod.\ Phys.\ {\bf#1}, #2 (#3)}
\def \rp#1{~~~~~\ldots\ldots{\rm rp~}{#1}~~~~~}
\def \yaf#1#2#3#4{Yad.\ Fiz.\ {\bf#1}, #2 (#3) [Sov.\ J.\ Nucl.\ Phys.\
{\bf #1}, #4 (#3)]}
\def \zhetf#1#2#3#4#5#6{Zh.\ Eksp.\ Teor.\ Fiz.\ {\bf #1}, #2 (#3) [Sov.\
Phys.\ - JETP {\bf #4}, #5 (#6)]}
\def \zpc#1#2#3{Z.\ Phys.\ C {\bf#1}, #2 (#3)}
\def \zpd#1#2#3{Z.\ Phys.\ D {\bf#1}, #2 (#3)}


\begin{thebibliography}{99}
\bibitem{better} A fit to the complete energy and angular dependence
(\ref{dist}) would be a more sensitive probe for the chirality than the
asymmmetry ${\cal A}$.
\end{thebibliography}
\end{document}